\documentstyle[draft]{l-aa}

\thesaurus{}

\newcommand{\etal}{{\it et al.}}
\newcommand{\half}{\mbox{$\frac{1}{2}$}}
\newcommand{\kms}{\mbox{\ km\ s$^{-1}$}}

\newcommand{\msun}{\mbox{$M_{\odot}\;$}}
\newcommand{\mstar}{\mbox{$M_{\ast}\;$}}
\newcommand{\rsun}{\mbox{$R_{\odot}\;$}}
\newcommand{\rstar}{\mbox{$R_{\ast}\;$}}

\newcommand{\ltappeq}{\mathrel{\hbox{\rlap{\hbox{\lower4pt\hbox{$\sim$}}}\hbox{$<$}}}}

\newcommand{\mnras}{{\sl MNRAS}}
\newcommand{\apj}{{\sl ApJ}}

\newcommand{\apjs}{{\sl ApJS}}

\newcommand{\aanda}{{\sl A\&A}}
\newcommand{\pasp}[1]{{\sl PASP} {\bf #1}}

\title{On the spin-down of Be stars}
\author{John M. Porter}
\institute{Astrophysics Research Institute,
School of Engineering,
Liverpool John Moores University,
Byrom Street, Liverpool L3 3AF, UK}

\date{Recieved March 6; accepted March 30, 1998}

\begin{document}
\maketitle

\begin{abstract}
The spin-down of Be stars due to angular momentum transport from star to
disc has been considered. This has been prompted by empirical studies
of observed optical and IR line profile studies indicating that the
disc is rotating in a Keplerian fashion.
It is found that substantial spin-down may occur, especially for late
B stars throughout their main-sequence lives for the ``strongest''
discs (most dense $\sim 10^{-11}$g cm$^{-3}$ with high radial velocity
$\sim 1\kms$ at their inner edge and with large opening angle $\sim
15^\circ$). This 
is in conflict with studies of rotational velocity distributions for
different luminosity classes, which show no significant evolution.
The implications of this for Be star discs are considered.


\end{abstract}

\section{Introduction}

Be stars are now widely accepted to have two distinct regions of circumstellar
matter : a diffuse polar stellar wind and a dense equatorial ``disc''
(Dachs 1987, Slettebak 1988).
One of the major objectives of Be star research is to develop a theory
which describes both of these components. The fast diffuse polar wind
is well described by standard line-driven wind theory 
(Castor, Abbott \& Klein 1975, Friend \& Abbott 1986, Kudritzki \etal\ 1989).
However, it has been much more difficult to describe the
equatorial disc. Empirical models of the disc structure have been
presented by e.g. Marlborough (1969), Waters (1986) and Hanuschik
(1996), whilst theoretically driven models have been developed by
Poe \& Friend (1986), Chen \& Marlborough (1992), Bjorkman \&
Cassinelli (1993),  Willson (1986), Ando (1986) and Lee, Saio \& Osaki
(1991).
These involve phenomena such as magnetic winds, latitudinal variation
of driving lines, wind compression, stellar
pulsation and viscous excretion.
The most promising disc theory for several years, Bjorkman \&
Cassinelli's (1993) wind compressed disc model, has been shown to be
incapable of reproducing observed discs 
by Owocki, Cranmer \& Gayley (1996) and Porter (1997) via
different routes.

One feature of empirical studies of line profiles in Be star discs is
that they imply a rotationally supported disc, and that the rotation
falls off in an Keplerian fashion (e.g. Dachs \etal\ 1986, Hanuschik
1989, 1996). The half-line width is typically larger
than $v {\rm sin}i$ (Hanuschik 1996).

Also, the current model for V/R variations (described in e.g. Dachs
1987) assumes that $m = 1$ perturbations arise in a Keplerian disc
(Papaloizou \etal\ 1992).
Excretion disc models proposed by Lee, Saio \& Osaki
(1991) provide naturally this sort of rotationally supported disc --
the disc material is rotating at its Keplerian speed, and drifts
outward due to the effect of viscosity. 
These models require that angular momentum is supplied
at the inner boundary of the disc. Given a prescription for the
viscosity (e.g. '$\alpha$' from Shakura \& Sunyaev 1973), then the
surface disc density, and disc scale height may be integrated from the
equations conserving angular momentum and mass (e.g. see Pringle 1981).
A similar model has been used by Pringle (1991) applied to the
cessation of accretion (``decretion'') by a young stellar object.

In this investigation, assuming that the disc is indeed rotationally
supported, the spin-down of the central star is calculated. 
In \S2 the evolution of a star's rotational velocity is derived
given that it is supplying angular momentum to the disc. Estimates of
spin-down times are presented in \S3 across the B star range of
stellar parameters. This is discussed in \S4 and conclusions are
presented in \S5.

\section{Angular momentum transfer \& spin down}

It is here assumed that the disc
material is moving round the star in approximately a Keplerian
fashion. The star, however is not rotating at its Keplerian velocity,
and so some angular momentum needs to be transferred to the
disc. This may be achieved through non-radial pulsations (described by Osaki
1986) or through the action of magnetic fields. However, the exact
mechanism is not important for the discussion here.

The specific angular momentum of the gas at the star's surface
$l_\ast$, rotating at velocity $v_{\phi, \ast}$ is 
\begin{equation}
l_\ast = v_{\phi,\ast} R_\ast = f
\left( \frac{2G\mstar}{3R_p} \right)^{\half} R_\ast,
\end{equation}
where $R_\ast$ and $R_p$ are the equatorial and polar radius of the
star respectively ($R_\ast > R_p$), \mstar is the mass of the star,
and the star is rotating at a fraction $f$ of its critical (or
break-up) velocity ($G$ is the gravitational constant). 

This should be compared to the specific Keplerian angular momentum
$l_k = v_{\phi, k} r = \left(G\mstar r\right)^{\half}$. 
It is now assumed that the extra angular momentum is added to the disc
at or close to the star's surface -- this is a conservative estimate
leading to a lower bound on the angular momentum deficit.
If the
disc has a half-opening angle $\theta$, and $\theta$ is small, then
the rate of angular momentum supplied to the disc is 
\begin{eqnarray}
\frac{dL}{dt} & =  & (l_k - l_\ast) 4\pi\rstar^{\!\!2}\theta{\cal F}
\nonumber \\ 
 & = & 4\pi\rstar^{\!\!2} \theta 
\left(G\mstar \rstar\right)^{\half} 
\left\{ 1 - f \left(\frac{2\rstar}{3R_p} \right)^{\half} \right\}, 
\end{eqnarray}
where ${\cal F}$ is the mass flux through the disc in g~cm$^{-2}$s$^{-1}$.

The rate of angular momentum lost by the star is then simply
$dL_\ast/dt = -dL/dt$. If the star rotates as a solid body at angular
velocity $\Omega_\ast$, and is regarded as a polytrope of index
3/2, then the total angular momentum of the star is
\begin{equation}
L_\ast = \epsilon \mstar\!\!R_p^2 \ \Omega_\ast = 
\epsilon f \mstar\!\!\!^\frac{3}{2} 
\left( \frac{2 G R_p }{ 3 } \right)^{\half}
\end{equation}
where $\epsilon = 0.2046$
(this is a slight underestimate as the star's mean radius has been
assumed to be $R_p$).
This leads to the rate of angular momentum loss by the star:
\begin{equation}
\frac{dL_\ast}{dt} = \epsilon \mstar\!\!\!^\frac{3}{2} \left(\frac{ 2
G R_p}{3} \right)^{\half} \frac{df}{dt}
\end{equation}
which may be combined with eq.2 to finally give
\begin{equation}
\frac{df}{dt} = -
\left( \frac{4\pi\theta}{\epsilon} \right)\!\!
\left( \frac{{\cal F}\rstar^{\!2}}{\mstar} \right)\!\!
\left( \frac{3 \rstar}{2R_p} \right)^{\!\half}\!\!
\left\{ 1 -  f \left(\frac{2\rstar}{3R_p} \right)^{\half} \right\}
\end{equation}

\subsection{Disc parameters}
To calculate the spin-down, the parameters of the
disc need to be fixed. 
Many of the models for Be star discs may be described via power law
expressions of density, disc height and velocity. 
For example, both Lee \etal's (1991) and Okazaki's (1997) excretion
disc models may be represented by power law density distributions. 

The discs must also be normalized in some way such that they provide
sufficient IR excess as the various disc models outlined above do not yield
absolute values of densities, mass-loss rates etc.
Here the density at the inner edge of the
disc is used from Waters, Cot\'{e} \& Lamers' (1987) fitting to IR continuum
excess. 

The mass-flux in Waters \etal's model is 
${\cal F} = 7.7\times 10^{-6} v_0 \rho_{-11} $
where the density $\rho_{-11}$ is measured in 10$^{-11}$g~cm$^{-3}$,
the velocity at the inner edge of the disc $v_0$ in \kms and 
${\cal F}$ in g~cm$^{-2}$s$^{-1}$.
Waters \etal\ calculate densities of $\rho_{-11}~\approx~0.05$--2
(their table 4) across a range of B spectral sub-types.
The velocity of the flow assumed by Waters \etal\ was 5\kms (yielding
a high mass flux) is almost certainly too high (Marlborough \etal\ 1997).
Exactly what value of the initial velocity should be used depends
upon which model is implemented, although $v_0$ has been calculated to
be subsonic from both empirical and theoretical models.
Poeckert \etal\ (1982), Marlborough \& Cowley (1974) \&
Marlborough \etal\ (1997) find initial velocities of $\ltappeq 1\kms$
using empirical models, whilst for excretion disc models 
it can be
shown that the radial velocity of an ``alpha'' accretion (or
excretion) disc is $v_r \sim \alpha c_s^2 / v_\phi$, where $c_s$ is the
sound speed (e.g. Pringle 1981) which becomes
\begin{equation}
v_0 \approx 0.3 \alpha T_4 \left( \frac{\rstar}{\mstar}
\right)^{\half} \; \kms
\end{equation}
where $T_4$ is the disc temperature in 10$^4$K, and \mstar and \rstar
are measured in Solar units. This leads to values for the velocity of
$v_0 = 0.2\alpha$--$0.6\alpha$\kms across the range of stellar
parameters for stars (see below).
The final unknown in eq.5 is the disc half-opening angle $\theta$.
This have been calculated statistically to be $\theta
\sim 5^\circ$ (Porter 1996) to $\theta \sim 13^\circ$ (Hanuschik 1996)
for Be star samples, whilst Waters \etal\ (1987) use an
opening angle of $\theta = 15^\circ$. 

With the above mass flux normalized to a base density, this should
ensure that the disc should be able to produce enough free-free
emission in the IR to account for the observed excesses. 
However, this is still subject to the functional form of the density
distribution and thickness of the disc model. The above normalization
is therefore a necessary, but not sufficient procedure for all disc
models. The study by Marlborough \etal\ (1997) illustrates the
effect of differing power law density distributions on the optical and
IR line profiles.

\begin{figure}
\begin{picture}(100,430)
\put(0,0){\includegraphics{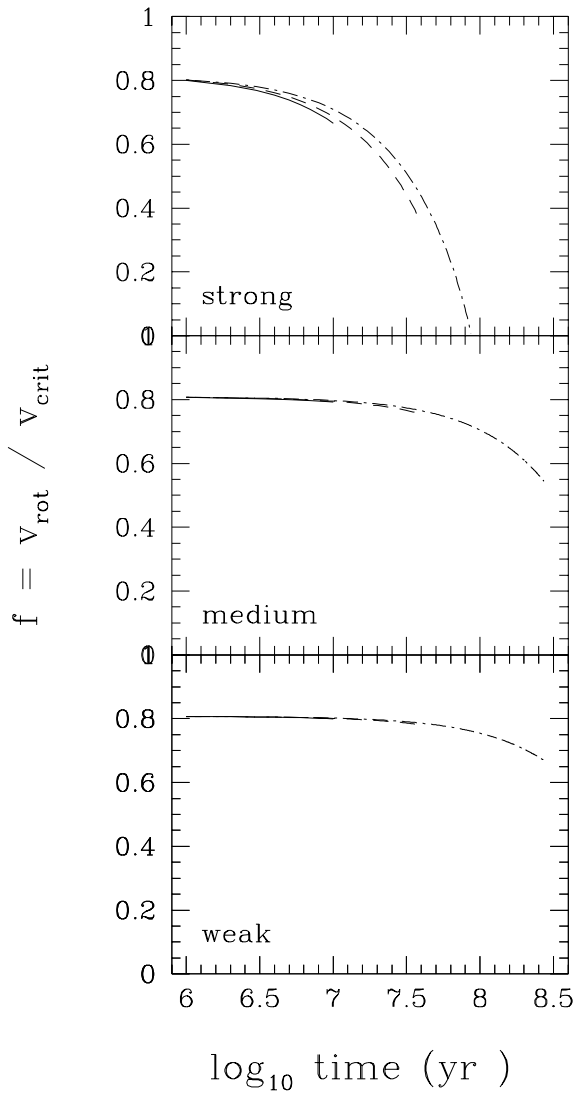}}
\end{picture}
\caption{The function $f(t)$ given by eq.7 for different disc models
and different stellar parameters. For each panel the solid,
dashed and dash-dotted lines correspond to the 17.5\msun,
7.7\msun and 3.4\msun models.
An initial value of $f = 0.81$ at $t = 0$ is assumed.}
\end{figure}

\section{Results}
Combining the estimate for mass flux with eq.5 gives
\begin{eqnarray}
\frac{df}{dt} & = & -1.4\times 10^{-15} 
\left ( \frac{R_p^{2}}{\mstar} \right )
\left( v_0 \theta \rho_{-11} \right) \times \nonumber \\
& & \ \ \ \ \ \ \ \ \ \ \ \ \ \ \ \ \ \ \ \ 
\left( \frac{\rstar}{R_p} \right)^{\!5/2}\!\!
\left\{ 1 -  f \left(\frac{2\rstar}{3R_p} \right)^{\half} \right\}
\end{eqnarray}
where $R_p$ is measured in Solar units.
This function may now be integrated to obtain the evolution of the
rotation of a Be star. Three model stars (with
stellar parameters in Table 1) spanning the B spectral type are used.

Eq.7 breaks into three separate parts: the stellar parameters, the
disc parameters, and a function of rotation.
For a given star the models may be characterised by the 
parameter $P = v_0\theta\rho_{-11}$.
With $\theta~=~5$--$10^\circ$, $\rho_{-11}~=~0.05-2.0$, and
$v_0~=~0.1$--$1.0$\kms, the disc parameter ranges between 
$P~=~(0.5$--$350)\times10^{-3}$.   
Three values are used; $P = 5\times 10^{-3}$,
$10^{-2}$, and  $10^{-1}$ corresponding to ``weak'', ``medium'' and
``strong'' discs. 

Figure 1 displays the spin-down due to each of these discs with the
top, middle and bottom panels corresponding to the strong, medium and
weak discs respectively. The initial value of $f$ was set to 0.81, and
the calculation for each star was stopped at a time equal to the main
sequence lifetime. This was taken to be the time at which hydrogen
burning ceases in the models of Maeder \& Meynet (1989). These times
are given in column 4 of table 1.

\begin{table}[t]
\caption{Stellar parameters for the stars considered from
Schmidt-Kaler (1982). The radii quoted are the polar radii. $'\tau$ is
an approximation to the main sequence lifetime taken from the hydrogen
burning times from Maeder \& Meynet (1989)}
\begin{tabular}{cccc}
Spectral & \mstar  & \rstar  & $\tau$ \\
type     & (\msun)   &  (\rsun) & (yr) \\ \hline
 B0      & 17.5    &  7.7 &  $9.8\times 10^6$ \\
 B3      & 7.7     &  4.7 &  $3.9\times 10^7$ \\
 B9      & 3.4     &  2.8 &  $2.7\times 10^8$
\end{tabular}
\end{table}

The results displayed in fig.1 clearly illustrate three things: 
(i) the amount of spin-down is strongly dependent on the disc
parameter $P$,
(ii) the amount of spin-down for a star is not strongly dependent on
its stellar parameters, except through its lifetime, and
(iii) for some combinations of disc and star, substantial
spin-down will occur.

For early B stars, there is only significant spin-down during the main
sequence for strong discs. Both medium and weak discs
cause little evolution in the rotation. However,
for late B stars unless $P < 5\times 10^{-3}$
significant spin-down should
occur. If there is a threshold rotational velocity below which no disc
should form (previous studies have found this to be $f \sim 0.2-0.3$;
Porter 1996), then late B stars may even spin-down so much that they
cease to support a disc at all.

\section{Discussion}

It has been demonstrated above that there is a possibility of some Be stars
significantly spinning down.
However, in observational studies, Fukuda (1982), Slettebak (1982) 
and more
recently Zorec \& Briot (1997) have found that across the B star
range, there is no significant change in the rotational velocity
distributions between different luminosity classes.
This, contrasted with the above numerical results makes it very
difficult to see how, (for late B stars), even medium discs may be
generated, as these stars should have spun-down significantly during
their main sequence lives, or with strong discs they should have
spun-down to small rotational velocities and may cease to support a
disc. 
As the disc structure has little or no dependance on the spectral type
of the star (van Kerkwijk \etal\ 1995) then, at a first glance, it is
perplexing that observational studies indicate that the discs 
around Be stars rotate at Keplerian velocities.

It is possible that all the actual Be star discs are what have
been termed weak here. Some small fraction of stars may indeed have strong 
discs and even though they may spin-down significantly during their
lifetime they may not skew the rotational distributions enough to
detect. The conclusion of this is that the disc population is
dominated by weak discs. 
The least well known parameter in the disc is
the initial radial velocity $v_0$. Assuming maximal values of
$\rho_{-11}~\approx~2$ and $\theta~=~15^\circ$ for the disc,
then a star will not spin down significantly if the initial
radial velocity is $v_0~\ltappeq0.01$\kms ($P~\ltappeq~5\times 10^{-3}$).

The estimates of $v_0$ stem from optical and IR line
modelling of discs although these only provide upper limits of $\sim
1\kms$. 
Excretion disc models (e.g. Lee \etal\ 1991) do predict values of $v_0$,
although at the expense of introducing the viscosity parameter
$\alpha$ (eq.6). With $\alpha \ltappeq 0.01$, values of $v_0$
applicable to weak discs are obtained.
Imposition of no evolution in the rotational distribution therefore
may be seen as a tighter constraint on $v_0$ (or on $\alpha$).

In the discussion so far, the disc has been assumed to exist around
the star at all times in the main-sequence lifetime of the
star. Be stars are known to undergo phases when the disc is apparently
lost (see e.g. Hirata 1995 and references therein). Depending on the
fraction of their 
lifetime spent without discs (or with very weak discs), Be stars could
easily support strong discs and still not evolve significantly in
rotational velocity. 
 
It is timely to discuss whether the disc material is
genuinely being lost. Strong discs could exist around B stars if all,
or a fraction of them are re-accreted at some phase and hence
spinning-up the B star. Indeed, a similar process of viscous
re-accretion of disc
material has been suggested by Hanuschik \etal\ (1993) in their study
of $\mu$~Cen. 
However, at low radial velocities such as these it is impossible to
detect whether material is inflowing or outflowing as the rotation
dominates the kinematics in the disc (also see Hanuschik \etal\
1993). This renders the line profiles
generated in the disc impotent as radial velocity detectors in the
presence of the Keplerian rotation. Consequently it seems impossible
to obtain
an explicit observational measurement of $v_0$.

\section{Conclusion}

Under the assumption that Be stars maintain discs throughout their
main-sequence lives, 
spin-down is found to occur for certain parameters of the
circumstellar disc if the disc rotates in a Keplerian fashion. 
This is in conflict with observational studies of the distribution of
rotational velocity for Be stars of different luminosity classes. 
The simplest resolution to this paradox is that the majority of Be
star discs must occupy parameter space in which
$P~=~v_0\theta\rho_{-11} \ltappeq 5\times 10^{-3}$. 
For the range of opening angles from observational studies
(e.g. Porter 1996), and densities (e.g. Waters \etal\ 1987) this
implies very low radial velocities of $v_0~\ltappeq~0.01$\kms.

\section*{Acknowledgements}
I. Negueruela \& I. Steele are thanked for their careful reading and
constructive criticism of the first draft of this paper. The referee,
H. Henrichs, is thanked for constructive criticism of the manuscript.
JMP is supported by a PPARC postdoctoral research assistantship.

{}

\end{document}